\documentstyle[aps,multicol,epsfig]{revtex}

\begin{document}
\draft
\title{Spin dynamics and ordering of a cuprate stripe-antiferromagnet}
\author{G.B. Teitel'baum\cite{GBT}, I.M. Abu-Shiekah, O. Bakharev and H.B. Brom}
\address{Kamerlingh Onnes Laboratory, Leiden University,P.O.Box 9504, 2300 RA\\
Leiden, The Netherlands}
\author{J. Zaanen}
\address{Instituut Lorentz for Theoretical Physics, Leiden\\
University, P.O. Box 9506, 2300 RA Leiden, The Netherlands}
\date{\today}
\maketitle

\begin{abstract}
In La$_{1.48}$Nd$_{0.4}$Sr$_{0.12}$CuO$_{4}$ the $^{139}$La and $^{63}$Cu NQR relaxation rates and
signal wipe-out upon lowering temperature are shown to be due to purely magnetic fluctuations. They
follow the same renormalized classical behavior as seen in neutron data, when the electronic spins
order in stripes, with a small spread in spin stiffness (15\% spread in activation energy). The La
signal, which reappears at low temperatures, is magnetically broadened and experiences additional
wipe-out due to slowing down of the Nd fluctuations.
\end{abstract}


\begin{multicols}{2}
\settowidth{\columnwidth}{aaaaaaaaaaaaaaaaaaaaaaaaaaaaaaaaaaaaaaaaaaaaaaaaa}

Strongly correlated electron systems such as layered cuprates exhibit very unusual properties. One
of the most interesting among them is the coexistence of superconductivity with local
antiferromagnetism (AF) -- a fingerprint of the topological effects of doping of AF insulators by
holes. The charges segregate into a periodical array of stripes separating antiphase
antiferromagnetic domains. Experimental evidence for stripe correlations has been provided by
neutron studies in Nd-doped La$_{1.875}$Sr$_{0.125}$CuO$_{4}$ and in other cuprates and
nickelates\cite {Tranquada95,Tranquada99}. The spatial organization of the stripe structures is a
subject of much debate\cite {Zaanen98,Roepke99,Nachumi98,Hunt99,Curro00,Suh00}. Stripe formation is
characterized by the temperatures of charge ($T_{{\rm charge}}$) and spin $T_{{\rm spin}}$)
ordering with $T_{{\rm charge}}>T_{{\rm spin}}$. Since these different types of order coexist on
the microscopic level, local methods of analysis, like NMR/NQR, are well suited to see their
interrelation. One striking feature in the NMR data is the wipe-out effect. In Cu-NQR experiments
on a number of Sr doped La$_{2}$CuO$_{4}$ samples, Imai {\it et al.}\cite{Hunt99} showed a
correlation between the amount of the intensity loss and the development of charge order of the
stripe phase. Curro {\it et al.}\cite{Curro00} found strong Cu wipe-out effect in their NMR
experiments on La$_{2-y-x}$Eu$_{y}$Sr$_{x}$CuO$_{4}$ and showed that this effect could be accounted
for by a wide (100\%) distribution in the energy of the thermally activated correlation times that
determine the relaxation processes - so called glassy behavior.

In this Letter we resolve this apparent controversy. We show that wipe-out effects are a beautiful
consequence of the growing spin order in the stripe phase, by taking profit of the NQR frequency
range of $^{139}$La and $^{63}$Cu, and especially of the low frequencies and relatively small line
widths of La NQR in La$_{1.48}$Nd$_{0.4}$Sr$_{0.12}$CuO$_{4}$. In this compound both Cu and La
exhibit strong wipe-out effects. Because La (contrary to Cu) nuclei are relatively weakly coupled
to the electronic spins in the CuO$_2$ planes, La NQR signals can be followed down to the spin
ordering temperature, as seen by $\mu$SR. Using the spin correlation times extracted from the
activated La spin-lattice relaxation rates we are able to predict precisely these wipe-out features
by introducing a spread of only 15\% in the activation energy. Within experimental error this
energy agrees with the value found from neutron data, where the relation with stripe ordering was
well established\cite{Tranquada99}, and is explained in the renormalized classical model. An
additional finding is the reappearance of a magnetically broadened NQR signal at low temperatures.
For the 6 MHz La transition the recovery is maximal around 4~K, where still about half of the La
nuclei are missing.

Experimentally we measure the $T$ dependence of the signal intensities $\tilde{I}$\cite{intensity}
and of the relaxation rates of the three $^{139}$La NQR transitions ($I$=7/2) at 6, 12 and 18 MHz
for La$_{1.48}$Nd$_{0.4}$Sr$_{0.12}$CuO$_{4}$ and those of $^{63,65}$Cu ($I$=3/2) around 36 MHz.
The question whether spin or charge fluctuations are relevant is answered by comparison of the
rates of the $^{63}$Cu and $^{65}$Cu and precisely monitoring the magnetization recovery curves
after spin reversal for the various La transitions. All relaxation rates are purely due to spin
fluctuations. Knowing that the fluctuations are magnetic, we extend the approach of Hammel {\it et
al.}\cite{Curro00,Suh00} to obtain the proper analytic description of  the wipe-out effect. With a
simple signal visibility criterium and the known values of the hyperfine couplings, from the
wipe-out curves correlation times for the spin dynamics are calculated. At the end we show that the
La linewidth increase below 20 K is due to the internal hyperfine field induced by the ordered Cu
moments and that Nd fluctuations are responsible for the missing La NQR signal intensity at the
lowest temperatures.

Let us now discuss our findings in more detail. NQR measurements were performed on a powder sample
\cite{Teitelbaum98}. The preparation is described in ref.\cite{Buchner91}. Susceptibility $\chi$
measurements at 0.001~T show a superconducting transition temperature of 5~K. The intensity
$\tilde{I}$ multiplied by $T$ and corrected for $T_2$ is shown in Fig.1. Because the nuclear
magnetization follows a Curie law, $T\tilde{I}$ is expected to be $T$ independent. This relation is
not obeyed, see Fig.1. Instead, $\tilde{I}T$ strongly decreases with decreasing $T$, the so-called
wipe-out being different for Cu and La. In Fig.1 arrows indicate the charge ($T_c \sim 65$~K) and
spin order ($T_s \sim 54$~K) temperatures as seen by neutrons\cite{Tranquada95}, and the magnetic
transition seen by $\mu$SR ($T_m \sim 31$~K)\cite{Nachumi98}. The LTO-LTT transition is around
68~K.
\begin{figure}[htb]
\begin{center}
\leavevmode \epsfig{figure=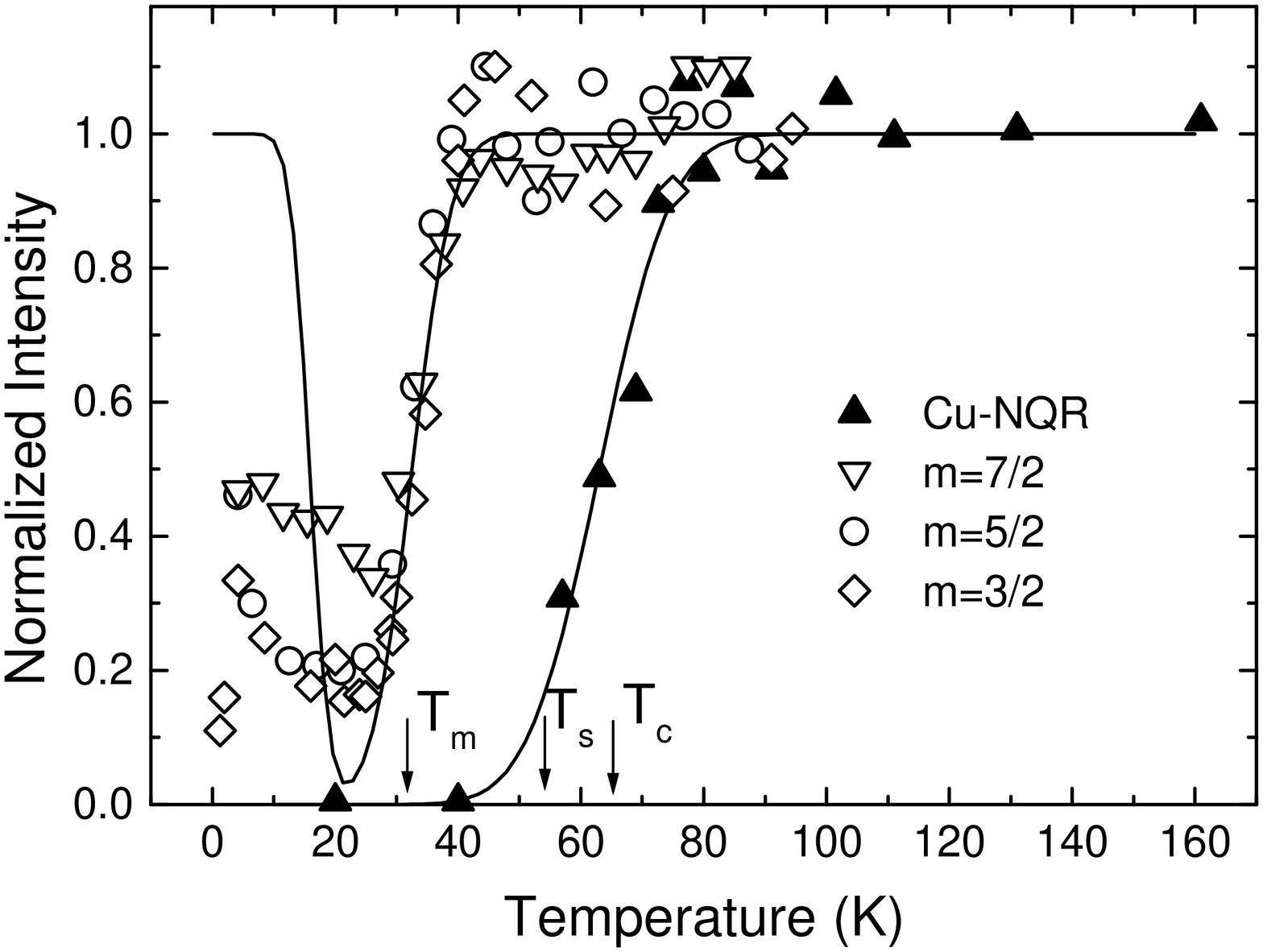,height=7cm,angle=-90}
\end{center}
\caption{Wipe-out in La and Cu NQR. For both Cu isotopes wipe out starts around 70~K, while for the
3 satellites of $^{139}$La this temperature is around 40~K. Drawn lines are predictions from the
model discussed in the text.} \label{f1}
\end{figure}

\begin{figure}[htb]
 \begin{center}
\epsfig{figure=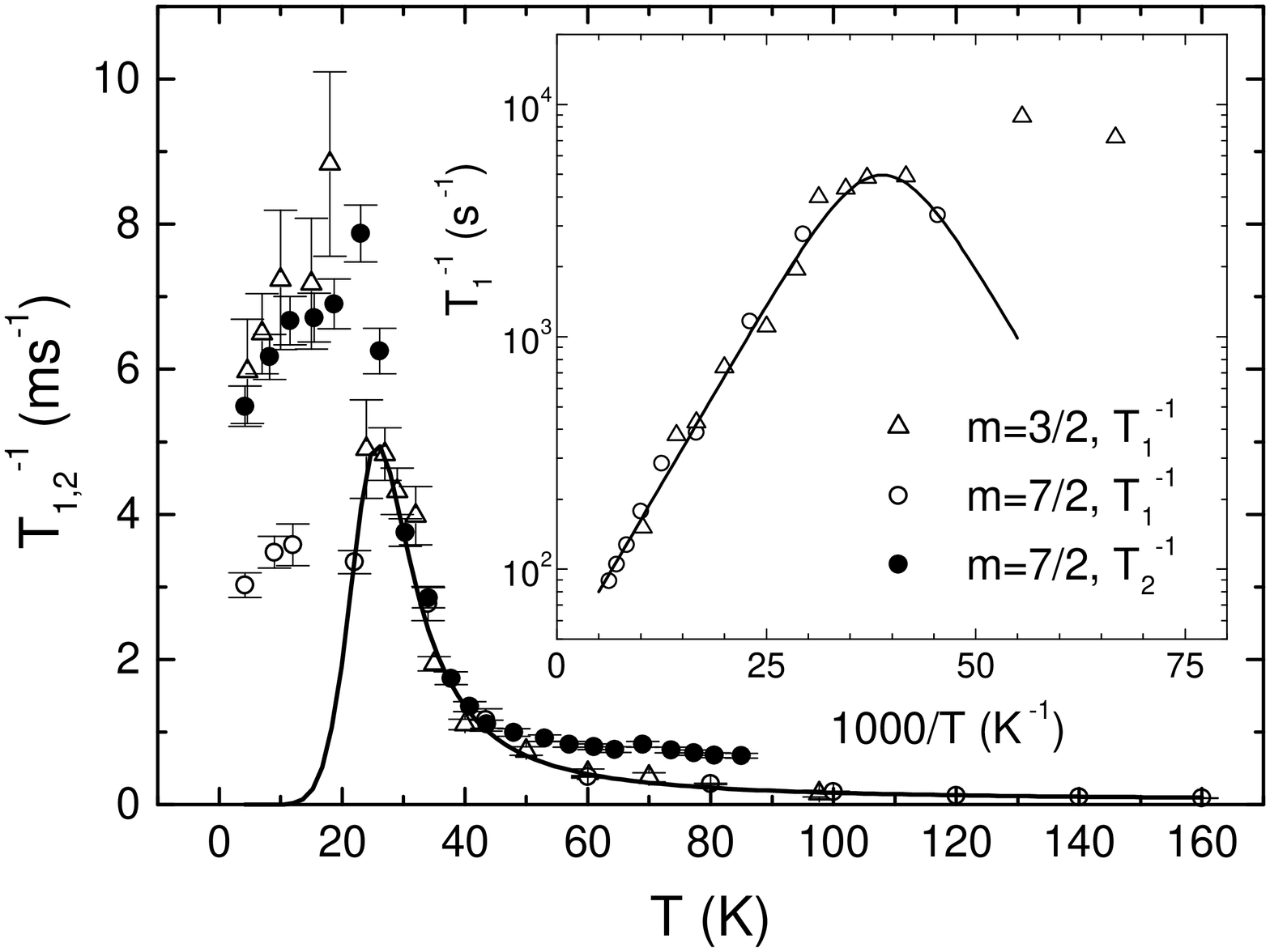,height=7cm,angle=-90} \caption{$^{139}$La $T_1^{-1}$ and $T_2^{-1}$ as
function of $T$.
 The solid line
 is a fit based on activated behavior with $E_0=143$~K. The deviations below
 20~K are due to the magnetic ordering.} \end{center}
\label{f2}
\end{figure}

The spin-lattice ($T_1^{-1}$) and spin-spin ($T_2^{-1}$) relaxation rates for the several
La-quadrupolar transitions, Fig.2, peak around 20~K, where also the wipe-out has its maximum. The
labels $m$ = 7/2, 5/2, and 3/2 refer resp. to the ($\pm 7/2,\pm 5/2$), ($\pm 5/2,\pm 3/2$), and
($\pm 3/2,\pm 1/2$)\cite{note1} transitions. Fits are made with stretched exponentials ([1-
$M(t)$]-recovery is $\propto \exp-(t/T_1)^{\alpha}$), indicating the presence of a distribution in
rates; the more $\alpha$ deviates from 1 the larger the influence of the distribution is. Here
$\alpha$ decreases almost linearly from 1 at 300~K to 0.6 at 20~K. Down to 30~K the $T$ dependence
can be described by $T_{1}^{-1} =W^2 \tau/(1+\omega^2\tau^2)$\cite {Slichter91} (characteristic for
exponential time correlation between fluctuating electronic spins) with $\tau = \tau_{\infty}
\exp(E/k_BT)$, $W$ a matrix element and $E$ an activation energy, see drawn line in Fig.2. The $T$
dependence of $T_2$, see Fig.2, is determined by the same activation law. From the fit we obtain $E
=143 \pm 5$~K. With the known hyperfine coupling \cite{Goto94}, we estimate $\tau_{\infty}$ as
$4\cdot 10^{-12}$~s. The same value is found from the maximum in $T_1^{-1}$.

To see whether the relaxation processes are determined by magnetic or electric fluctuations, we
compared the rates for $^{63}$Cu and $^{65}$Cu. The Cu rates at 71~K were 8.1 ($^{63}$Cu) and 9.8
ms$^{-1}$ ($^{65}$Cu) and at 130~K resp. 7.3 and 10.1 ms$^{-1}$. If $\omega \tau \ll 1$,
$T_{1}^{-1}$ is $\ $proportional to $W^{2}\tau $. For the magnetic case the ratio of the $^{63}$Cu
and $^{65}$Cu transition rates is proportional to $(\gamma_{63}/\gamma _{65})^{2}=0.87$, while in
case of electric transitions it is the ratio between the quadrupolar moments squared, which equals
1.14. The found ratio's show Cu relaxation to be magnetic. For La only the rates for the various
quadrupolar transitions are available. Here we make use of the fact that the fundamental transition
probability, that appears in the exponents of the relaxation expression \cite{Watanabe94} is
weighted by well defined factors, that are different for magnetic or electric processes. At 130~K,
60~K, 33~K, 28~K and 4.2~K, the magnetization recovery curves after application of a $\pi $ pulse
follow stretched exponentials with rates that were a factor $1.8 \pm 0.15$ faster for $m=5/2$ than
for $m=7/2$. This value agrees with the magnetic ratio of 1.9.

How to explain the pronounced wipe-out features? Imai {\it et al.}\cite {Hunt99} suggested that the
intensity loss might be related directly or indirectly\cite{noteind} to the growth of the stripe
order parameter with decreasing temperature. Let us restrict ourselves to the direct case and to
simplify the argument, suppose that the fluctuating stripe order leads to random jumps between the
two NQR frequencies which correspond to the extremal values of charge distribution and differ by a
value $\delta \omega $. The signal decay for $\delta \omega \tau _{{\rm ch}}\ll 1$ will be given by
the $\exp \{-t[1/T_{2}+(\delta \omega \tau _{{\rm ch}})^{2}/(8\tau _{{\rm ch} })]\}
$\cite{Slichter91}. The resulting decay is not only determined by the standard magnetic term, but
also by the dephasing due to the electric fluctuations (life time $\tau _{{\rm ch}}$). Since the
relaxation rates are governed by magnetic fluctuations, charge fluctuations can at most weakly
contribute to the wipe-out phenomena\cite{notec}.

More generally, wipe-out effects have been shown to be linked to charge/spin fluctuations having a
distribution $P(E)$ in activation energies $E$ and hence in correlation times\cite{note2,Curro00}.
In case of a gaussian distribution of $E$, the extrapolated intensity of the signal at t=0
($\tilde{I}(0)$) is given by: $\tilde{I}(0)=(1/{\sqrt{2\pi }\Delta } )\int\limits_{0}^{\infty }\exp
(-{(E-E_{0})^{2}}/{2\Delta ^{2}})dE$, with $E_0$ the mean activation energy, and $\Delta$ the width
of the distribution. In the echo pulse sequence $\pi /2$-$t_{r}$-$\pi $-$t_{r}$ the delay time
$t_{r}$ allows a registration in the echo of only those nuclei, that do not relax too fast.
Therefore only part of nuclei will contribute in $\tilde{I}$ leading to a special cut-off of the
integration. Let us assume that we are only seeing those nuclei of which the signal has decayed by
a factor of $f$ or less at time $2t_{r}$ - i.e. for which $1/{T_{2R}}=({\Omega ^{2}\tau })/({
1+\omega ^{2}\tau ^{2}})\leq A$. For magnetic fluctuating fields $\Omega ^{2}=\beta \gamma
^{2}{h_{0}}^{2}$, $A=(\ln f)/2t_{r}$, $h_{0}$ denotes the hyperfine field probed by the nuclei and
$\beta =(2+r)/3$\cite {Stern95} with the anisotropy factor $r=3.6$ for Cu and $\beta = 6$ for La
(as deduced from our own relaxation data). The boundary values of $\tau $ follow from $A\omega
^{2}\tau ^{2}-$ $\Omega ^{2}\tau $ $+$ $A=0$ and are given by $\tau _{1,2}=({\Omega ^{2}\pm
\sqrt{{\Omega ^{4}}-4A^{2}\omega ^{2}}})/{2A\omega ^{2}}$. With $\tau =\tau _{\infty }\exp
(E/k_{B}T)$\cite {note2}, and hence $E_{i}=k_{B}T\ln (\tau _{i}/\tau _{\infty })$, the expression
for the intensity at $t=2t_{r}$ becomes
\begin{equation}
{\tilde{I}}_{2t_{r}}(0)\propto \int\limits_{0}^{E_{2}}e^{-\frac{(E-E_{0})^{2} }{2\Delta
^{2}}}dE+\int\limits_{E_{1}}^{\infty }e^{-\frac{(E-E_{0})^{2}} {2\Delta ^{2}}}dE  \label{eq3}
\end{equation}
being proportional to the quantity of nuclei influenced by the magnetic fluctuations with the
lifetimes beyond the interval between $\tau _{1}$ and $\tau _{2}$ \ ( the integrals are directly
related to the error functions). In Eq.(\ref{eq3}) $\tilde{I}$ is corrected for an exponential loss
factor $\exp (-2t_{r}/T_{2})$. There appear two bands in the solution, which contribute to the
signal: a band of high-frequency fluctuations (smaller activation energies $E<E_{2}$) and a band of
low-frequency fluctuations (larger activation energies $E>E_{1})$. The values of $E_{1}$ and
$E_{2}$ are linear functions of $T$ and the gap $E_{1}-E_{2}=k_{B}T\ln (\tau _{1}/\tau _{2})$
between them is the NMR wipe-out gap. The condition for the gap to exist is very simple $\Omega
^{2}>2A\omega $. The presence of two bands, see Eq.(\ref{eq3}), gives rise to the reentrant
behavior of the echo-amplitude with lowering $T$.

In case of La NQR, $\Omega _{{\rm La}}$ is rather small, and the condition for the wipe-out gap is
realized for $\tau $ lying in the narrow interval around $\tau ={\Omega ^{2}}/{2A\omega ^{2}}$.
Using $A_{{\rm La}}\sim 10^{5}$ ~s$^{-1}$ ($f\sim e^{5}$ and $t_{r}=30$ $\mu $s), the $^{139}$La
hyperfine coupling constant (1.7 kOe/$\mu _{{\rm B}}$)\cite{Goto94} and $\Omega _{{\rm La}}\sim
6\cdot 10^{6}$ s$^{-1}$ we obtain that for this interval the typical fluctuation times are $\tau
\sim $ 10$^{\text{-8 }}$s. For the Cu nuclei $\Omega ^{2}/A\omega ^{2}\leq \tau \leq A/\Omega
^{2}$. With $A_{{\rm Cu}}\sim A_{{\rm La}}\sim 10^{5}$ s$^{-1}$, the $^{63}$Cu hyperfine coupling
constant of 139 kOe/$\mu _{{\rm B}}$, and $\Omega _{{\rm Cu}}\sim 6\cdot 10^{8}$~s$^{-1}$, it
follows that the wipe-out at 75 K is due to the fluctuations with $\tau \sim 10^{-11}-10^{-12}$~s.
Neglecting effects of the magnetic ordering of Cu (and the Nd) moments, the reappearance of the Cu
NQR signal will take place for extremely slow fluctuations with $\tau \sim 10^{-6}$~s, realized
only at very low temperatures.

The drawn lines in Fig.1 are fits to the wipe-out behavior with the numerical constants calculated
above. The free parameters are in principle $E_0$, $\Delta$, and $\ln(\tau_i/\tau_{\infty})$. If
for $E_0$ the same value is used as for the relaxation data, i.e. $E_0 = 143 \pm 5$~K, the fit to
the Cu and La data gives mutually consistent values for the other free parameters: $\Delta = 21 \pm
3 $~K and $\tau_{\infty}$ equals the value found from the relaxation data. Note that for the low
frequency La transitions the wipe-out is more pronounced, since the wipe-out gap is $\propto
1/\omega ^2$.

An activated $T$ dependence of $\tau $ can have many causes. However, a most natural interpretation
is in terms of the behavior of the relaxation time of a classical quasi 2D Heisenberg
antiferromagnet which is on its way to its 3D phase transition. The relaxation time is set by the
magnetic correlation length $\xi $\cite{Tye89} and the latter behaves like $\xi (T)=e^{T^{\ast
}/T}/(2T^{\ast }+T)$ where $T^{\ast }=2\pi \rho _{s}$ in terms of the spin-stiffness $\rho _{s}$.
According to our relaxation and wipe out data $T^{\ast }=143\pm 5$~K which is consistent with the
$T^{\ast }=200\pm 50$~K as deduced by Tranquada et al\cite{Tranquada99} from the $T$ dependence of
$\xi $ as measured by neutron scattering. This spin stiffness associated with the stripe
antiferromagnet is an order of magnitude smaller than the one of the pure antiferromagnet of
half-filling. If the spin system would be classical the implication would be that the exchange
interactions mediated by the charge stripes would be smaller by two orders of magnitude as compared
to the exchange interaction inside the magnetic domains. This is inconsistent with the persistence
of anti-phase correlations up to rather high energies as seen by inelastic neutrons scattering.
Moreover, there is no doubt that the spin system is highly quantum-mechanical at short length
scales and the stripe antiferromagnet should exhibit {\em renormalized} classical
behavior\cite{Chakravarty89}. This implies that the spin system should be in the proximity of a
quantum-phase transition to a disordered state and it is well understood that the renormalized
stiffness diminishes when this transition is approached, while the spin velocity is barely
changing. Hence, the small spin-stiffness of the stripe antiferromagnet signals that this system is
much closer to the quantum phase transition than the half-filled antiferromagnet, in agreement with
theoretical expectations \cite{vDuin98,Sachdev00}.

To evaluate the role of the Nd ion on the correlation times, we also determined the relaxation
rates in La$_{1.71}$Eu$_{0.17}$Sr$_{0.12}$CuO$_4$\cite{Hunt99,Suh00,Teitelbaum99}. The $^{139}$La
relaxation rates were about a factor 10 lower than in the 0.4Nd compound. With the hyperfine
coefficients used in 0.4Nd, the correlation times for the fluctuating fields derived from
$T_1^{-1}$($^{63}$Cu) and $T_1^{-1}$($^{139}$La) above 20~K were the same. The different values of
$\tau $ reflect that Nd and Eu ions induce a different pinning strength for stripes in the LTT
phase, whereas the equal hyperfine constants show that above 20~K  Nd does not influence the La
nuclear relaxation rates directly.

To determine whether spins or charges are responsible for the final La line shapes, we have
followed the line profiles for various satellites (due to its small splitting, $m=3/2$ is the most
sensitive) as function of $T$, see Fig.3. Above the wipe-out regime the La line widths scale with
their splitting, which show them to be electric. Below 20~K the linewidths increase due to the
presence of an internal magnetic field, see Fig.3. The drawn line represents the mean field
staggered magnetization for $S=1/2$. The saturated value of the additional width (full width at
half intensity), obtained by taking the square root of the difference in second moments of the
broadened and unbroadened line, amounts to 2 .0 MHz, close to the splitting seen in undoped
La$_{2}$CuO$_{4}$, where the $m=3/2$ splitting is 2.5 MHz \cite{MacLaughlin94}. In the undoped
compound (with an ordered moment in the Neel state of $\sim 0.55 \mu_B$) the splitting can be
reproduced by a field of 0.11~T perpendicular to the electric field gradient (with anisotropy
parameter $\eta =0.02$ and the usual in-plane field angle $\phi=0$). Here the saturated splitting
seen for the $m=3/2$ line can be simulated by an external field of 0.08~T, again applied
perpendicular to the electric field gradient ($\eta = 0.13$ is fixed by the line positions above
the magnetic ordering and $\phi=\pi/4$). As (see below) Nd moments are not yet involved, we
estimate the Cu ordered moment in 0.4Nd to be $\sim 0.4\mu_B$. The missing spectral weight of about
50\% at 4.2~K for $m=3/2$(La) might be explained by an internal field of the same order as the
quadrupolar splitting of 6 MHz felt by the unseen La-sites. Such a scenario agrees with the $\mu
$SR finding \cite{Nachumi98} that most or all $\mu $SR sites are magnetic. However, 0.1~T (2 MHz)
at 4.2~K is about the maximum field at the La sites (even with
Nd\cite{Tranquada99,Roepke99,Teitelbaum98}) one might expect. The $T$ dependence of
$\tilde{I}(m=3/2)$ below 4.2~K shows that we deal with additional wipe-out caused by slow Nd-spin
fluctuations. This extra channel in $T_2^{-1}$(La) becomes important close to the ordering
temperature of 1~K of the Nd moments\cite{Teitelbaum98} and partially destroys the recovery of
echo-signal predicted by Eq.(1).
\begin{figure}[htb]
\begin{center}
\leavevmode \epsfig{figure=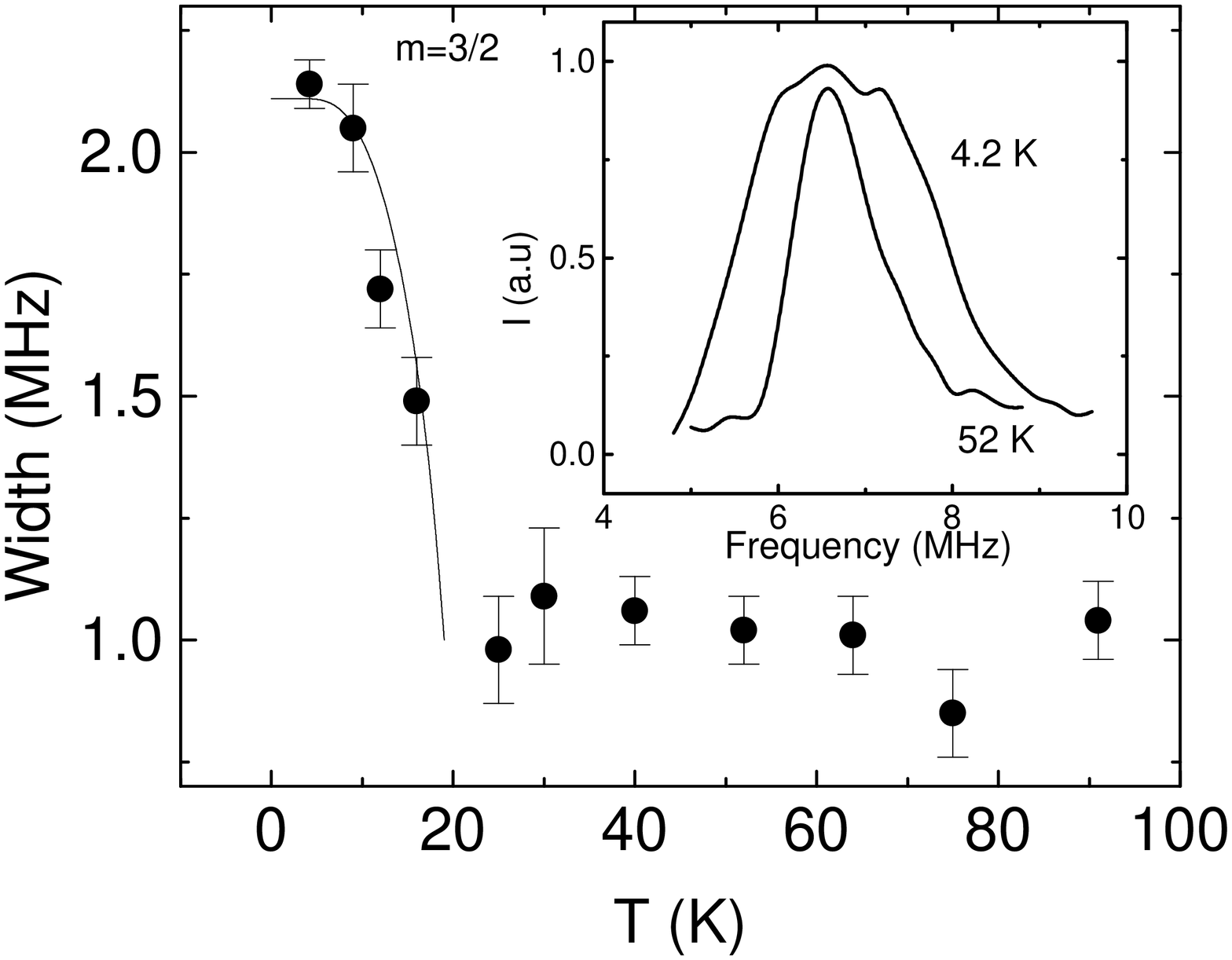,height=7cm,angle=-90}
\end{center}
\caption{$T$ dependence of the $^{139}$La linewidth for $m=3/2$. Magnetic ordering sets in below
20~K (drawn line: mean field fit). The inset shows the changes in the line profile.} \label{f3}
\end{figure}

In summary, the wipe-out features of Cu and La in the temperature regime above the spin ordering
transition find a natural explanation in terms of the well understood fluctuations of a
quantum-antiferromagnet which is approaching its ordered state. However, for the 0.4Nd compound
this `ordered' state is not straightforward as wipe-out persists down to 1~K for the majority of La
spins. Here La wipe-out proceeds in two stages, of which the first is due to slowing down of Cu
spins and the second below 4~K is dominated by fluctuations of Nd magnetic moments.

This work is supported in part by the Dutch Science Foundation FOM-NWO and by the State HTSC
Program of the Russian Ministry of Sciences (Grant no. 98001) and by the Russian Foundation for
Basic Research (Grant no. 98-02-16528). O.G.A. Berfelo is acknowledged for his assistance in the
measurements.


\end{multicols}


\begin{references}
\bibitem{intensity}  For $\tilde{I}$(La) the whole line is integrated and if
needed corrected for sensitivity changes with $\omega $. $\tilde{I}$(Cu) refers to peak values of
the line and is therefore sensitive for changes in lineshape, which occur below 20~K.
\bibitem{note1}  In case of magnetic order the $\pm 1/2$ state will no
longer be pure, see Ref.\cite{MacLaughlin94}.
\bibitem{noteind} Charge ordering in stripes will not only slow down charge fluctuations but might
also change the magnetic fluctuation spectrum considerably inducing wipe-out, see
\protect\cite{Hunt99}.
\bibitem{notec}  The additional decay is the strongest for $\delta \omega
\tau _{{\rm ch}}\sim 1$, when the echo signal is already suppressed by magnetic fluctuations.
\bibitem{note2}  The activated behavior is suggested by the relaxation data
and biased by theory (see last part).
\end{references}
\end{document}